\documentclass[acus]{epac98}
\usepackage{epsfig}

\title{HALO FORMATION IN SPHEROIDAL BUNCHES WITH SELF-CONSISTENT 
STATIONARY DISTRIBUTIONS}
\author{A.V.~Fedotov, R.L.~Gluckstern, University of Maryland, 
College Park, MD 20742, USA \\
\underline{S.S.~Kurennoy}, R.D.~Ryne, Los Alamos National Laboratory,
        Los Alamos, NM 87545, USA }

\begin{document}
\maketitle

\begin{abstract} 
A new class of self-consistent 6-D phase space stationary 
distributions is constructed both analytically and numerically. 
The beam is then mismatched longitudinally and/or transversely, 
and we explore the beam stability and halo formation for the 
case of 3-D axisymmetric beam bunches using particle-in-cell 
simulations. We concentrate on beams with bunch length-to-width 
ratios varying from 1 to 5, which covers the typical range of 
the APT linac parameters. We find that the longitudinal halo 
forms first for comparable longitudinal and transverse mismatches. 
An interesting coupling phenomenon --- a longitudinal or transverse 
halo is observed even for very small mismatches if the mismatch 
in the other plane is large --- is discovered.
\end{abstract} 

\section{Introduction}

High-intensity applications of ion linacs, such as the transformation 
of radioactive waste, the tritium production \cite{APT}, and drivers 
for spallation neutron sources \cite{SNS}, require peak beam 
currents up to 100~mA with final energies about 1 GeV and beam 
losses below 1~ppm. 
Understanding mechanisms of intense-beam losses, in particular, beam 
instabilities and halo formation, is of primary importance to 
satisfy these stringent requirements.

Most efforts in halo formation study have been concentrated so far 
on 2-D (and often axisymmetric, essentially 1-D) beams, 
see \cite{Reiser} and references therein. 
While it produced some analytical results for the simplest 
case, the K-V distribution, for more realistic 
distributions particle-core model and particle-in-cell (PIC) 
simulations have been used, \cite{RLG94}-\cite{RLG&SK}. 
As was recognized from these studies, an rms mismatch of the beam to 
the focusing channel is the main cause of the halo formation.

To single out and explore the mechanism of halo formation associated 
with the beam rms mismatch, it is important to start from an initial 
distribution that satisfies the Vlasov-Maxwell equations and, 
therefore, remains stationary for the matched case.
A beam with some initial {\it non} stationary distribution will evolve 
from its initial state even being rms-matched to the channel, due to 
redistribution effects (its evolution is caused by mismatches in 
higher moments). For 2-D axisymmetric beams, a set of stationary 
distributions with a sharp beam edge was constructed and
explored in \cite{RLG&SK}: 
\begin{equation}
  f_n(H) = \left \{ \begin{array}{cc}
       N_n n (H_0-H)^{n-1} &  \mbox{for } H \leq H_0 \ ,  \\
                   0 &  \mbox{for } H > H_0 \ ,  \label{fn}
                \end{array} \right.
\end{equation}
where $H$ is the hamiltonian of the transverse motion,
$H_0=const$, and $N_n$ are normalization constants.
The set includes the K-V distribution as a formal limit of $n \to 0$, 
as well as more realistic ones, like waterbag ($n=1$) and other 
distributions, with higher non-linearities in space-charge forces. 
In this paper, we present results of a similar program in the 3-D 
case. More details can be found in \cite{HaloSt}.

\section{Stationary 3-D Distribution}
 
\subsection{Analytical Consideration}
 
We consider a smoothed external focusing with gradients $k_z$, $k_y$, 
$k_x$. In general, the beam bunch can be chosen to have an 
approximately ellipsoidal boundary. For simplicity, we concentrate 
on the axisymmetric case $(k_x = k_y)$, for which the bunch is 
approximately spheroidal. 
Our axisymmetric 6-D phase space distribution is
\begin{eqnarray}
f(\mbox{\boldmath $R$},\mbox{\boldmath $p$}) 
 = N(H_0 - H)^{-1/2} \ , \quad  \mbox{ where}    \label{f3d} \\
H = k_xr^2/2 + k_zz^2/2 + 
e\Phi_{sc}(\mbox{\boldmath $R$}) + mv^2/2 \ .  \label{H3d}
\end{eqnarray}
Here $\mbox{\boldmath $p$} = m\mbox{\boldmath $v$}$, $r^2 = x^2 + y^2$, 
and $\Phi_{sc}(\mbox{\boldmath $R$})$ is the electrostatic
potential due to the space charge. We work in the bunch Lorentz frame,
where all motion is non-relativistic. 

The distribution (\ref{f3d}) is analogous to (\ref{fn}) with $n=1/2$. 
Since all its dependence on the coordinates is through the
hamiltonian $H=H(\mbox{\boldmath $R$},\mbox{\boldmath $p$})$, which is
an integral of motion, the distribution is stationary. 
The same would be true for other exponents in (\ref{f3d});
however, for the particular case of -1/2, the Poisson equation in
3-D case is linear. Namely, it can be written as
\begin{equation}
\nabla^2 G(\mbox{\boldmath $R$}) = -k_s + 
 \kappa^2 G(\mbox{\boldmath $R$}),       \label{Geq}
\end{equation}
where $k_s = 2k_x + k_z$, 
$\kappa^2 = (eQ/\epsilon_0)/\int d\mbox{\boldmath $R$} 
G(\mbox{\boldmath $R$})$, $Q$ is the bunch charge, and
\begin{equation}
G(\mbox{\boldmath $R$}) \equiv H_0 - k_xr^2/2 - 
k_zz^2/2 - e\Phi_{sc}(\mbox{\boldmath $R$}) \ . 
\end{equation}

The solution to Eq.~(\ref{Geq}) for a spheroidal shaped bunch can 
be written in the spherical coordinates $R$, $\theta$ 
($\cos \theta \ = z/R  ,\  \sin \theta = r/R$) as
$G(\mbox{\boldmath $R$}) = (k_s/\kappa^2) g(\mbox{\boldmath $R$})$, 
where
\begin{equation}
g(\mbox{\boldmath $R$}) = 1 + \sum^{\infty}_{\ell = 0} \alpha_{\ell}
P_{2\ell} (\cos \theta ) i_{2\ell} (\kappa R) \ .   \label{gser}
\end{equation}
Here $P_{2\ell} (\cos \theta )$ are the even Legendre
polynomials and $i_{2\ell} (\kappa R)$ are the spherical Bessel 
functions (regular at $\kappa R = 0$) of imaginary argument.
Since $g(\mbox{\boldmath $R$})$ is proportional to the charge 
density, the bunch edge is determined by the border 
$g(\mbox{\boldmath $x$}) = 0$, closest to the origin. 
We choose $\alpha_{\ell}$'s to approximate a spheroidal surface 
with semiaxis $a$ in the transverse direction and $c$ in the 
longitudinal one, $r^2/a^2 + z^2/c^2 = 1$.  

From the equations of motions, we express the rms tune 
depressions as 
\begin{equation}
\eta^2_{x,{\rm rms}} \equiv \frac{m\langle \dot{x}^2\rangle}
{k_x\langle x^2\rangle} \ , \ \eta^2_{z,{\rm rms}} \equiv 
\frac{m\langle \dot{z}^2\rangle}{k_z\langle z^2\rangle} \ .
\end{equation}
Note also that $m\langle \dot{x}^2\rangle = m\langle \dot{y}^2\rangle
=m\langle \dot{z}^2\rangle = m\langle v^2\rangle/3$, because $H$ 
depends only on $v^2$ and $\mbox{\boldmath $R$}$.  Thus our choice 
of the form $f(H)$ {\em automatically} corresponds to equipartition
(equal average kinetic energy in the three spatial directions). 
The values of $\alpha_{\ell}$ in Eq. (\ref{gser}) for given
$c/a$ and $\kappa a$ are found by minimizing
$\oint ds g^2(\mbox{\boldmath $R$})$ along the boundary. 
For a fixed bunch shape $c/a$, the rms tune depressions depend 
on the dimensionless parameter $\kappa a$ (see in \cite{HaloSt}).
A contour plot of $g(\mbox{\boldmath $R$})$ for
a typical case $c/a = 3$, $\kappa a = 3.0$ is shown in Fig.~1. 
This range of parameters corresponds to the Accelerator Production 
of Tritium (APT) project \cite{APT}.  

\begin{figure}[htb]
\centerline{\epsfig{figure=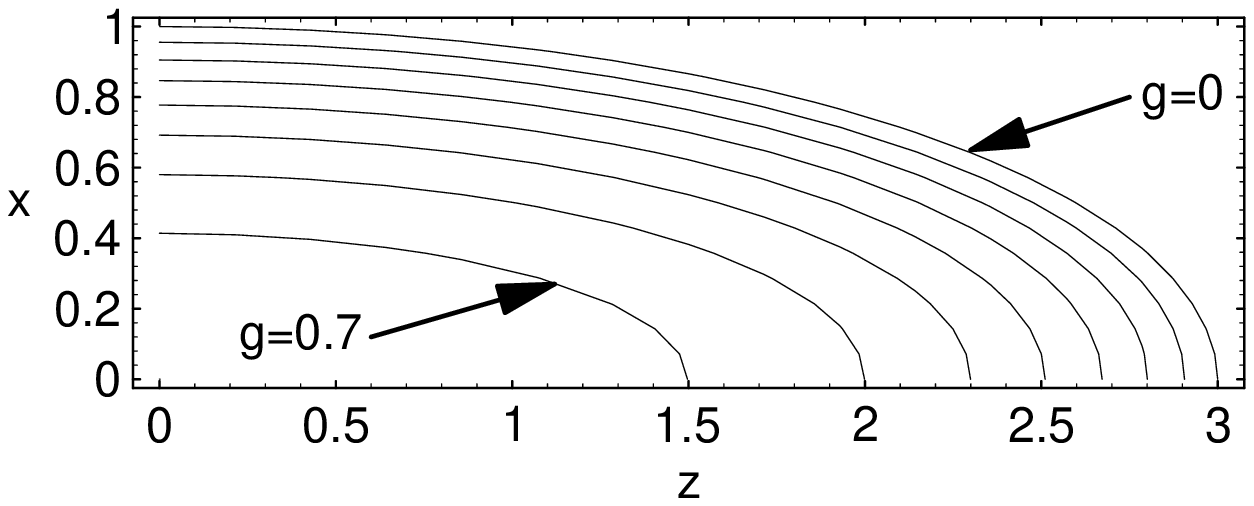,width=82.5mm}}
\caption{Charge density contours $g(\mbox{\boldmath $R$}) = const$ 
for $c/a=3$, $\eta_x = 0.65$, $\eta_z = 0.49$.}
\end{figure}

\subsection{Numerical Investigation}

A 3-D particle-in-cell (PIC) code has been developed to test the 
analytic model of normal modes \cite{HaloSt} in the distribution 
Eq. (\ref{f3d}) and to explore halo formation. 
The single-particle equations of motion are integrated 
using a symplectic, split-operator technique.  The space charge 
calculation uses area weighting (``Cloud-in-Cell'') and implements 
open boundary conditions with the Hockney convolution algorithm. 
The code runs on parallel computers (we mostly used T3E machine at 
NERSC), and in particular, the space charge calculation has been 
optimized for parallel platforms. Up to $2.5 \cdot 10^7$ particles 
have been used in our simulation runs, with $10^6$ being a typical 
number. 

Initially, the 6-D phase space is populated according to 
Eq.~(\ref{f3d}), and then the $x,y,z$ coordinates are mismatched 
by factors $\mu_x = \mu_y = 1 + \delta a/a$, $\mu_z = 1 + \delta c/c$ 
and the corresponding momenta by $1/\mu_x = 1/\mu_y$, $1/\mu_z$. 
Simulations show that an initially matched distribution remains 
stable even for very strong space charge. Introducing some initial
mismatch leads to the oscillations of the core, and later on the
beam halo develops, as shown in Fig.~2. This figure shows maximal 
values $z_{max}$ and $x_{max}$ of the longitudinal and transverse 
coordinates (in units of $a$) of the bunch particles versus time, 
for the case $\mu_x = \mu_z=\mu$. The jumps of $z_{max}$ and $x_{max}$ 
correspond to the halo formation moments; after that the distribution 
stabilizes. One can see that the longitudinal halo develops earlier 
than the transverse one for equal mismatches in both directions.
This is in accordance with our expectations since the longitudinal 
tune depression is lower for an elongated bunch.

\begin{figure}[htb]
\centerline{\epsfig{figure=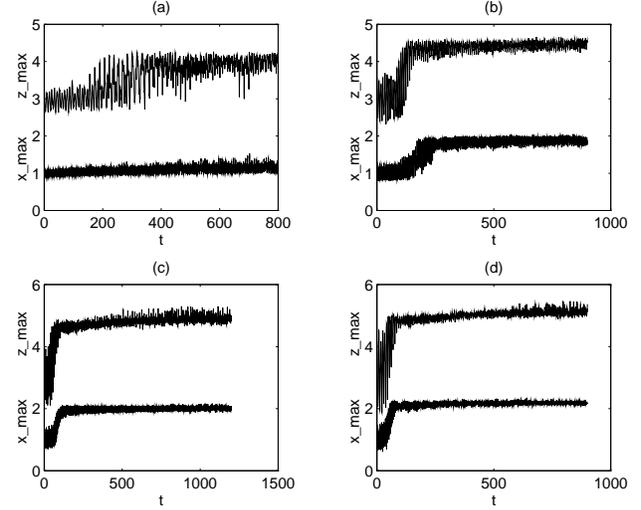,width=82.5mm}}
\caption{Halo development for increasing mismatches: a) $\mu = 1.1$, 
b) $\mu = 1.2$, c) $\mu = 1.3$, d) $\mu =1.4$. Time $t$ is in 
arbitrary units, $c/a=3$, $\eta_x=0.65$, $\eta_z=0.49$.}
\end{figure}

Choosing larger mismatch either longitudinally or
transversely, one can observe primarily the longitudinal 
or transverse halo, respectively. Results of a systematic study
for different bunch shapes $c/a$ and mismatch parameters are 
summarized below, first for the longitudinal case. 

We define the halo extent as a ratio of the halo maximal size to
that of a matched distribution. The {\bf longitudinal halo} extent 
is found to be approximately linearly proportional to the mismatch.
In addition, the ratio $z_{max}/(\mu c)$ slightly increases for
stronger space charge, from 1.2--1.3 for $\eta_z$ above 0.5 to
1.4--1.5 for $\eta_z < 0.4$. 
The halo intensity, defined roughly as the fraction of particles 
outside the bunch core, was also found depending primarily on the 
mismatch. Large mismatches (40\% and higher) lead to several percent 
of the particles in the halo, which is clearly outside acceptable 
limits for high-current machines. Obviously, serious efforts should 
be made to match the beam to the channel as accurately as possible.

For a fixed mismatch, the halo starts to develop earlier for more 
severe tune depression. Another interesting observation is that 
for purely longitudinal mismatches ($\mu_x=1$) in elongated bunches
($c/a>2$) the longitudinal halo intensity shows a strong dependence on
the mismatch. The number of particles in the halo drops dramatically 
with $\mu_z > 1$ decreasing; in fact, we see no halo for $\mu_z < 1.2$.
A similar threshold behavior was observed in 2-D case \cite{RLG&SK}.

The extent of the {\bf transverse halo} has a similar linear 
dependence on the mismatch: $x_{max}/(\mu a)$ depends weakly
on $\eta_x$, just slightly increasing from 1.4--1.5 for $\eta_x$ 
around 0.8 to 1.6--1.8 for $\eta_x < 0.4$. Again, the halo
intensity is governed primarily by the mismatch. 
In general, the transverse halo closely duplicates all the features 
observed for non-linear stationary distributions in 2-D simulations 
\cite{RLG&SK}. The only two differences seen are related to the 
moment and rate of halo development: first, in 3-D simulations it 
clearly starts earlier for severe tune depression, which was not the 
case in 2-D; and second, the transverse halo in the 3-D case develops 
significantly faster than in 2-D for comparable mismatches and tune 
depressions.  

Our 3-D simulations clearly show the {\bf coupling} between the 
longitudinal and transverse motion: a transverse or longitudinal
halo is observed even for a very small mismatch (less than 10\%)
as long as there is a significant  mismatch in the other plane.  
For example, in Fig.~3 we see a longitudinal halo for only 5\% 
longitudinal mismatch, when $\mu_x = \mu_y = 1.5$.
The coupling effect is noticeable even for modest mismatches.  
We mentioned above that $\mu_z \ge 1.2$ is required to observe a 
longitudinal halo when $\mu_x = 1$. However, when there is a mismatch 
in all directions, the halo develops even for $\mu_z = \mu_x = 
\mu_y = 1.1$ (10\% mismatch in all directions).
Such a behavior clearly shows the importance of the coupling effect.

\begin{figure}[htb]
\centerline{\epsfig{figure=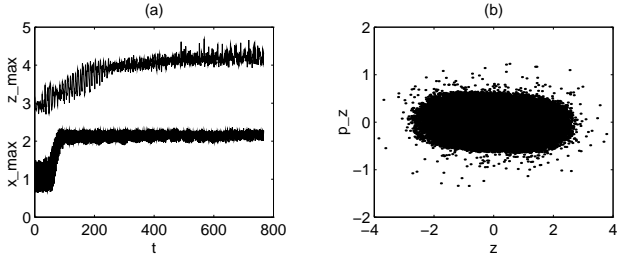,width=82.5mm}}
\caption{Coupling effect for $c/a = 3$, $\mu_x = \mu_y = 1.5$, 
$\mu_z = 1.05$: (a) maximum $x$ and $z$ versus time;  
(b) $z$-$p_z$ phase space diagram (plotted only 32K particles 
out of $10^6$ used in simulations).}
\end{figure}

\section{Summary and Discussion}

Unlike previous studies of 2-D models of long beams,
this paper addresses the beam stability and halo formation in a 
bunched beam with the parameters in the range of new high-current
linac projects \cite{APT,SNS}. A new class of 6-D phase space 
stationary distributions for a beam bunch in the shape of a prolate 
spheroid has been constructed, analytically and numerically. Our 
choice of parameters automatically assures equipartition. 
We therefore study the halo development in 3-D bunches which are 
in thermal equilibrium, without masking effects of the initial-state
redistribution. Such an approach allows us to investigate the major 
mechanism of halo formation associated with the beam mismatch.

Using our PIC code with smoothed linear external focusing forces, by
introducing an initial mismatch in the transverse and/or longitudinal 
directions we find that both transverse and longitudinal halos 
can develop, depending on the values of tune depressions and 
mismatches. 
An interesting new result is that, due to the coupling between the $r$ 
and $z$ planes, a transverse or longitudinal halo is observed for a 
mismatch less than 10\% if the mismatch in the other plane is large.
Our main conclusion is that the longitudinal halo is of great 
importance because it develops earlier than the transverse one for 
elongated bunches with comparable mismatches in both planes.
In addition, its control could be challenging. 
This conclusion agrees with the results \cite{Bar&Lund}
from the particle-core model in spherical bunches.

We expect only small quantitative differences for distributions 
(\ref{f3d}) with other exponents, not -1/2, 
based on results for the set (\ref{fn}) in 2-D \cite{RLG&SK}. 
More interesting are 3-D effects due to the phase-space 
redistribution of an initial non-stationary state. Our preliminary 
results from PIC simulations show that the redistribution process
can produce the beam halo in the same fashion as a small rms 
mismatch \cite{DifDistr}. A similar conclusion was made for 2-D 
axysymmetric beams \cite{Jam,Okam}. In 3-D, however, the effect 
can be amplified by the coupling, especially noticeable in the 
bunches with $c/a$ close to 1.   

The authors would like to acknowledge support from the U.S.\ 
Department of Energy, and to thank R.A.~Jameson and T.P.~Wangler 
for useful discussions.


\begin{thebibliography}{99}
\bibitem{APT} 
APT Conceptual Design Report, LA-UR-97-1329, Los Ala\-mos, NM, 1997.
\bibitem{SNS} 
SNS Conceptual Design Report, NSNS-CDR-2/V1, Oak Ridge, TN, 1997.
\bibitem{Reiser} 
M. Reiser, Theory and Design of Charged Particle Beams, Wiley,
New York (1994).
\bibitem{RLG94} 
R.L. Gluckstern, Phys. Rev. Lett. {\bf 73}, 1247 (1994).
\bibitem{Jam} 
R.A. Jameson, in `Frontiers of Accelerator Technology', 
World Scient., Singapore, 1996, p. 530.
\bibitem{SYL} 
S.Y. Lee and A. Riabko, Phys. Rev. E {\bf 51}, 1609 (1995).
\bibitem{TW} 
T.P. Wangler, et al, in Proceed. of LINAC96, 
Geneva, Switzerland (1996). - CERN 96-07, p.372.
\bibitem{RLGetal96} 
R.L. Gluckstern, W-H. Cheng, S.S. Kurennoy and H. Ye, Phys Rev. 
E {\bf 54}, 6788 (1996).
\bibitem{Okam} 
H. Okamoto and M. Ikegami, Phys. Rev. E {\bf 55}, 4694 (1997).
\bibitem{RLG&SK} 
R.L. Gluckstern and S.S. Kurennoy, in Proceed. of PAC97, 
Vancouver, BC, Canada (1997).
\bibitem{HaloSt}
R.L. Gluckstern, A.V. Fedotov, S.S. Kurennoy and R.D. Ryne, 
Univ. of Maryland, Phys. Dept. preprint 98-107, College Park,
MD, 1998; submitted to Phys. Rev. E.
\bibitem{Bar&Lund} 
J.J. Barnard and S.M. Lund, I \& II, in Proceed. of PAC97, 
Vancouver, BC, Canada (1997).
\bibitem{DifDistr}
A.V. Fedotov, R.L. Gluckstern, S.S. Kurennoy and R.D. Ryne, 
Univ. of Maryland, Phys. Dept. preprint 98-108, College Park,
MD, 1998; to be published.
\end{thebibliography}
\end{document}